%% file: main.tex
\documentclass[preprint,12pt,authoryear]{elsarticle}

\usepackage{amsmath}
\usepackage{amssymb}
\usepackage{bm}
\usepackage{graphicx}
\usepackage{booktabs}

\journal{Journal of High Energy Astrophysics}

\newcommand{\tgamma}{t_{\gamma,\mathrm{onset}}}
\newcommand{\dd}{\mathrm{d}}

\DeclareRobustCommand{\matern}{%
  \ifmmode
    \text{Mat\'ern-3/2}%
  \else
    Mat\'ern-3/2%
  \fi
}

\begin{document}

\begin{frontmatter}

\title{%
Nonstationary Stochastic Timing Signatures in the Prompt
Gamma-Ray Light Curve of GRB 170817A%
}

\author[ynu]{Lin Xie}

\author[ynu]{Dahai Yan\corref{cor1}}
\ead{yandahai@ynu.edu.cn}

\cortext[cor1]{Corresponding author}

\affiliation[ynu]{%
  organization={School of Physics and Astronomy, Yunnan University},
  city={Kunming},
  state={Yunnan},
  country={China}%
}

\begin{abstract}
We investigate time-dependent stochastic structure in the weak prompt gamma-ray emission of gamma-ray burst (GRB) 170817A using change-point and deep-kernel Gaussian-process (GP) models. The analysis is based on the 10--300 keV light curve observed by the Fermi Gamma-ray Burst Monitor (GBM) with 0.10 s time resolution. Two change-point configurations identify numerically similar covariance transitions at 0.269 and 0.237 s after the gravitational-wave merger, with fitted 10--90\% transition widths of 0.441 and 0.393 s, respectively. At the representative gate-defined boundary of $t_{\rm tr}=0.27$ s, all four tested fixed-split assignments yield positive evidence gains over the full-exposure stationary Mat\'ern-3/2 reference. The largest gain is $\Delta\ln Z_{\rm split}=5.72\pm0.16$ for the Mat\'ern-3/2$\rightarrow$Mat\'ern-3/2 assignment, while the gate-matched damped random walk (DRW)$\rightarrow$Mat\'ern-3/2 assignment gives $\Delta\ln Z_{\rm split}=5.50\pm0.16$. The fixed-split comparison therefore supports segment-specific covariance evolution without requiring a change of covariance family. The deep-kernel models recover localized time-deformation features peaking at 1.65 and 1.75 s after the merger for DRW and Mat\'ern-3/2 base kernels, respectively. Their offsets from the adopted gamma-ray onset at 1.74 s are $-0.09$ and $+0.01$ s, both within the 0.10 s sampling resolution. The full-band timing features remain stable when the bin width is changed from 0.10 to 0.12 s, whereas energy subdivision produces substantially larger shifts in the gate locations than in the warp peaks. Residual diagnostics show that the models reproduce the dominant temporal structure, although localized residual dependence remains. We interpret the gate-defined covariance transition and the localized time-deformation feature as model-dependent phenomenological timing diagnostics. Further simulation calibration and count-level modelling are needed to assess the statistical robustness and physical origin of the recovered nonstationary structure.
\end{abstract}


\end{frontmatter}

\section{Introduction}

The binary neutron-star merger GW170817 and its associated short gamma-ray burst (GRB)~170817A established the first direct connection between gravitational-wave (GW) emission and prompt high-energy radiation from a compact-binary merger \citep{Abbott2017PRL,Abbott2017ApJL,Goldstein2017,Savchenko2017}. This event has provided a unique reference for investigating the formation of relativistic outflows, the geometry of short-gamma-ray-burst jets, and the physical origin of prompt and afterglow emission. Its low apparent luminosity and unusual temporal and spectral properties
have motivated interpretations involving structured or off-axis relativistic
jets \citep{Lazzati2018,Mooley2018,Ghirlanda2019}, as well as cocoon and
shock-breakout emission \citep{Kasliwal2017,Gottlieb2018}.

Previous studies of the prompt emission of GRB~170817A have mainly focused
on its delay relative to the merger, duration, spectral evolution, energetics,
and pulse morphology
\citep{Goldstein2017,Savchenko2017,Abbott2017ApJL}. These observables provide
important constraints on the launching and propagation of the outflow.
In particular, the GW--gamma-ray delay can contain contributions from
central-engine activity and jet propagation through the merger ejecta
\citep{Geng2019,Zhang2019Delay}. The light curve can, however, also be regarded as a realization of an underlying stochastic process. From this perspective, its covariance structure characterizes the amplitude, correlation timescale, and temporal regularity of the variability, providing information complementary to conventional flux- and spectrum-based analyses.

Time-domain studies of high-energy variable sources commonly seek characteristic variability timescales and quasi-periodic oscillations (QPOs), but the significance of such features can be difficult to assess in the presence of stochastic red-noise variability. Gaussian processes (GPs) provide a flexible probabilistic framework for modelling correlated astronomical time series and for separating stochastic variability from more coherent temporal structure \citep{Rasmussen2006,Aigrain2023}. 

\citet{Yang2021} used GP stochastic-process models to reassess reported gamma-ray QPO candidates, showing that most variability could be explained by red noise, with only a limited number of QPO candidates remaining plausible. A complementary study by \citet{2021ApJ...919...58Z} identified a $\sim1.1$~yr QPO in PKS~0521--36 using GP modelling together with Lomb--Scargle, wavelet, and REDFIT analyses. These studies demonstrated the utility of GPs for distinguishing coherent temporal structure from stochastic red-noise variability. 

Subsequent work extended GP modelling from QPO searches to the characterization of stochastic variability itself. \citet{Zhang2022} used DRW and SHO processes to infer characteristic variability timescales in relativistic-jet sources, while \citet{Zhang2023} applied the approach to multiwavelength variability. \citet{Zhang2025} further showed that individual extreme flares can favour overdamped SHO or Mat\'ern-$3/2$ covariance structures over a simple DRW description. 

Most of these analyses, however, assume stationary covariance kernels within each analysed interval. This assumption can become restrictive for rapidly evolving transients whose variability amplitude or correlation structure changes on comparable timescales. \citet{2026ApJ...999..246Z} found that stationary DRW, SHO, and Mat\'ern-$3/2$ models can describe restricted flare intervals but may become inadequate when a larger fraction of the flare evolution is included, motivating change-point descriptions of evolving stochastic states. More generally, time-invariant covariance models may be insufficient for intrinsically nonstationary variability \citep{PaciorekSchervish2004}.

A departure from stationarity may indicate that different intervals of the light curve are characterized by different stochastic variability regimes, motivating GP constructions that allow covariance structure to vary across change points or more general change surfaces \citep{Herlands2016,Han2019}. Such a change need not coincide with an obvious discontinuity in the observed flux and may therefore be missed by analyses based only on pulse decomposition or time-resolved spectra \citep{Scargle_2013}. 

In this work, we characterize possible departures from stationary stochastic variability in the prompt gamma-ray emission of GRB~170817A. We use stationary GP models as reference descriptions and analyse the light curve with two complementary nonstationary constructions: a change-point model that represents an evolving covariance structure and a deep-kernel model that represents nonstationarity through a learned time deformation. We use these models to extract phenomenological timing signatures and evaluate their stability under changes in temporal binning and energy selection. For the gate-defined transition, we additionally perform a restricted Bayesian evidence comparison conditional on a fixed boundary inherited from the representative gate fit.

The paper is organized as follows. Section~\ref{sec:data} describes the gamma-ray data and timing conventions. Section~\ref{sec:method} introduces the stationary and nonstationary GP models and the model-comparison procedure. The results are presented in Section~\ref{sec:results}. Their possible astrophysical implications and the limitations of the analysis are discussed in Section~\ref{sec:discussion}, and our conclusions are summarized in Section~\ref{sec:conclusions}.

\section{Data and light-curve construction}
\label{sec:data}

We use publicly available time-tagged event (TTE) data from the
\textit{Fermi} Gamma-ray Burst Monitor (GBM)
\citep{Meegan2009} for GRB~170817A
\citep{Goldstein2017}. Events recorded by the NaI n1, n2, and n5 detectors are combined over the 10--300~keV energy range and binned at a uniform time resolution of $\Delta t=0.10$~s. The input time coordinate is already expressed in seconds relative to the GW170817 merger epoch. Throughout this work, $t=0$ corresponds to $T_{\rm GW}$, negative values of $t$ precede the merger, and positive values follow it. The background is estimated by fitting the count-rate evolution in off-source intervals and evaluating the resulting model over the analysis interval. The background-subtracted count rate in the
$i$th bin is then
\begin{equation}
y_i
=
R_{\rm obs}(t_i)-R_{\rm bg}(t_i).
\label{eq:background_subtraction}
\end{equation}
Assuming Poisson counting statistics, the rate uncertainty is approximated
as
\begin{equation}
\sigma_i
\simeq
\frac{\sqrt{C_{\rm obs}(t_i)}}{\Delta t},
\label{eq:rate_uncertainty}
\end{equation}
where $C_{\rm obs}(t_i)$ is the total observed count in the corresponding
time bin. Equation~\eqref{eq:rate_uncertainty} does not separately propagate
the uncertainty associated with the fitted background model.

The GP analysis uses the resulting time--rate--uncertainty array, $\{t_i,y_i,\sigma_i\}$, where $t_i$ is the input time in seconds relative to the GW170817 merger epoch. 
For numerical stability, the input relative-time coordinate is additionally
shifted according to
\begin{equation}
\widetilde{t}_i
=
t_i-t_{\min},
\qquad
t_{\min}
=
\min_i(t_i).
\label{eq:relative_model_time}
\end{equation}
The shifted coordinate $\widetilde{t}$ is used only internally during model
evaluation. It does not redefine the physical time origin.

\section{GP methodology}
\label{sec:method}

\subsection{GP likelihood}
\label{sec:gp_likelihood}

GP regression models the observed light curve as a random function with a specified mean function and covariance structure. For the binned count-rate measurements, the shifted model-time
coordinate $\widetilde{t}$ defined in
Equation~\eqref{eq:relative_model_time} is further standardized as
\begin{equation}
x_i =
\frac{\widetilde t_i-\bar{t}}
{s_t},
\label{eq:model_time}
\end{equation}
where $\bar{t}$ and
$s_t$ are the mean and standard deviation of the
shifted time samples, respectively. The standardized coordinate $x$
is used for the internal GP covariance calculations. The observed data vector is modelled as
\begin{equation}
\mathbf{y}\sim
\mathcal{N}
\left(
\boldsymbol{\mu},
\mathbf{K}
\right),
\label{eq:gp_distribution}
\end{equation}
where $\boldsymbol{\mu}$ is the mean function and $\mathbf{K}$ is the
covariance matrix evaluated at the model coordinates. In this work, we
adopt a constant mean function,
\begin{equation}
\mu(x)=\mu_0 ,
\label{eq:constant_mean}
\end{equation}
and construct the covariance matrix as 
\begin{equation} K_{ij} = k(x_i,x_j) + \left(\sigma_i^2+j_i^2\right)\delta_{ij}, 
\label{eq:covariance} 
\end{equation} 
where $k$ denotes the adopted covariance function, $\sigma_i$ is the measurement uncertainty of the $i$th binned count-rate measurement, $j_i$ denotes an additional white-noise (jitter) amplitude, and $\delta_{ij}$ is the Kronecker delta. For the stationary and deep-kernel models, the additional jitter is constant in time, $j_i=j$. For the change-point models, it is allowed to evolve with the same fitted gate.

The GP likelihood is then written as
\begin{equation}
\ln\mathcal{L}
=
-\frac12
(\mathbf{y}-\boldsymbol{\mu})^{\rm T}
\mathbf{K}^{-1}
(\mathbf{y}-\boldsymbol{\mu})
-\frac12\ln|\mathbf{K}|
-\frac{N}{2}\ln(2\pi),
\label{eq:gp_likelihood}
\end{equation}
where $N$ is the number of time bins. The likelihood provides the basis for all stationary and nonstationary GP models considered below.

\subsection{Stationary covariance models}
\label{sec:stationary_gp}

We use the DRW covariance function, equivalent to the Ornstein--Uhlenbeck (OU) covariance \citep{Rasmussen2006,Aigrain2023},
\begin{equation}
k_{\rm DRW}(z,z')
=
\sigma_{\rm DRW}^{2}
\exp\left(
-\frac{|z-z'|}{\tau_{\rm DRW}}
\right),
\label{eq:drw_kernel}
\end{equation}
and the Matérn-$3/2$ covariance function, hereafter denoted M32 in model labels and parameter or kernel subscripts,
\begin{equation}
\begin{aligned}
k_{\rm M32}(z,z')
&=
\sigma_{\rm M32}^{2}
\left(
1+\frac{\sqrt{3}|z-z'|}{\ell_{\rm M32}}
\right)\\
&\quad\times
\exp\left(
-\frac{\sqrt{3}|z-z'|}{\ell_{\rm M32}}
\right).
\end{aligned}
\label{eq:m32_kernel}
\end{equation}
Here $z$ denotes either the standardized model time $x$ or the learned
deep-kernel coordinate $u$. The parameters $\sigma_{\rm DRW}$ and
$\sigma_{\rm M32}$ determine the variability amplitudes, while
$\tau_{\rm DRW}$ and $\ell_{\rm M32}$ characterize the correlation scales.
The Matérn-$3/2$ process is smoother at short time separations than the DRW
process. These kernels are used as phenomenological descriptions of temporal
correlation.

The covariance parameters and additional white-noise amplitude are held constant in time. These models provide the full-exposure stationary reference descriptions against which the nonstationary alternatives are evaluated.

Because the kernels are fitted in the standardized coordinate, their physical
correlation scales are
\begin{equation}
\tau_{\rm phys}
=
s_t\,\tau_{\rm DRW},
\qquad
\ell_{\rm phys}
=
s_t\,\ell_{\rm M32}.
\label{eq:stationary_scale_conversion}
\end{equation}

\subsection{Change-point GP models}
\label{sec:change_point_gp}

To represent a transition between two covariance regimes, we refer to
\cite{Saatci2010GPChange} and define a time-dependent gate
\begin{equation}
q(x)
=
\left[
1+
\exp\left\{
-s(x-c)-\delta_{\boldsymbol{\phi}}(x)
\right\}
\right]^{-1}.
\label{eq:gate_function}
\end{equation}
The parameter $c$ specifies the centre of the analytic sigmoid component and
$s>0$ controls its sharpness. The correction
$\delta_{\boldsymbol{\phi}}(x)$ is generated by a two-layer feed-forward
network with Swish activations and is bounded according to
\begin{equation}
\delta_{\boldsymbol{\phi}}(x)
=
\delta_{\max}
\tanh\left[
g_{\boldsymbol{\phi}}(x)
\right],
\qquad
\delta_{\max}=0.10.
\label{eq:gate_residual}
\end{equation}
The bounded neural correction allows modest local departures from a
perfectly symmetric sigmoid, while the transition centre and overall
sharpness remain controlled primarily by the parametric gate.

The parametric gate centre in physical time is 
\begin{equation}
t_{\rm c}
=
t_{\min}
+
\bar{t}
+
s_t c.
\label{eq:gate_center}
\end{equation} 
When the gate is expressed on the physical-time coordinate, we write
$q(t)\equiv q[x(t)]$. For the analytic sigmoid component alone, $t_c$ corresponds exactly to the midpoint of the transition, i.e. $q(t_c)=0.5$. The bounded neural correction in Equation~\eqref{eq:gate_residual} allows only small local deviations from the analytic sigmoid and may therefore introduce a slight offset between $t_c$ and the numerical midpoint $t_{50}$ of the complete fitted gate. In the present fits, this correction is small and the difference between $t_c$ and $t_{50}$ is negligible for the timing interpretation. 

For completeness, the transition boundaries are determined from the complete sampled gate, including the neural correction. After fitting, we linearly interpolate between adjacent sampled points to obtain the physical times $t_{10}$, $t_{50}$, and $t_{90}$ at which $q(t)=0.1$, $0.5$, and $0.9$, respectively. The reported transition width is 
\begin{equation} \Delta t_{10-90}=t_{90}-t_{10}. 
\end{equation} 
We use $t_c$ as the representative transition centre throughout the paper, while $t_{10}$ and $t_{90}$ characterize the extent of the complete fitted transition profile.

The covariance structure before and after the transition is described by
independent zero-mean GP components with covariance functions
$k_{\rm pre}$ and $k_{\rm post}$, respectively. The gate function $q(x)$
smoothly interpolates between the two covariance regimes. The resulting covariance is
\begin{align}
k_{\rm cp}(x,x')
={}&
\left[1-q(x)\right]
\left[1-q(x')\right]
k_{\rm pre}(x,x')
\nonumber\\
&+
q(x)q(x')
k_{\rm post}(x,x').
\label{eq:change_point_kernel}
\end{align}
We consider both DRW-to-Matérn-$3/2$ and Matérn-$3/2$-to-DRW transitions, with independent
amplitudes and correlation scales for the two regimes. The gate construction is used here as a phenomenological interpolation between these covariance descriptions. In a closely related recent astrophysical application, a gated \matern{}-to-DRW covariance transition was used to localize the evolution of stochastic variability within a continuous X-ray exposure \citep{DongYan2026}.

The additional white-noise amplitude is allowed to evolve consistently with
the same gate:
\begin{equation}
j(x)
=
\left[1-q(x)\right]j_{\rm pre}
+
q(x)j_{\rm post}.
\label{eq:change_point_jitter}
\end{equation}

\subsection{Deep-kernel GP models}
\label{sec:deep_kernel_gp}

We adopt a deep-kernel covariance model in which a stationary base
kernel is evaluated in a learned monotonic temporal coordinate
\citep{Wilson2016},
\begin{equation}
k_{\rm deep}(x,x')
=
k_0
\left[
u(x),u(x');
\boldsymbol{\theta}
\right],
\label{eq:deep_kernel}
\end{equation}
where $x$ is the standardized model-time coordinate defined in
Equation~\eqref{eq:model_time}, $u(x)$ is a deterministic monotonic
transformation of that coordinate, $k_0$ denotes the stationary base
covariance kernel, and $\boldsymbol{\theta}$ contains its covariance
hyperparameters. We consider both the DRW and Matérn-$3/2$ kernels
defined in Section~\ref{sec:stationary_gp} as choices for $k_0$.

The base covariance is stationary with respect to separations in the
warped coordinate $u$, whereas the corresponding covariance expressed
in the original time coordinate is generally nonstationary because
equal intervals in $x$ need not map to equal intervals in $u$. Thus,
the transformation allows a stationary covariance family to represent
a smoothly varying effective correlation timescale in the observed
time domain. This construction follows the general deep-kernel
principle of learning an input representation jointly with the GP
covariance \citep{Wilson2016} and is closely related to input-warping
approaches for nonstationary GP \citep{2014arXiv1402.0929S}.

The warped coordinate is constructed from the ordered standardized
model-time samples $\{x_i\}$. Only the input time coordinate is
transformed; the observed count rates and their measurement
uncertainties remain those defined in Sections~\ref{sec:data} and
\ref{sec:gp_likelihood}. We first use a two-layer feed-forward neural
network with 32 hidden units per layer and Swish activations to define
a strictly positive auxiliary rate field,
\begin{equation}
r_{\boldsymbol{\phi}}(x)
=
{\rm softplus}
\left[
g_{\boldsymbol{\phi}}(x)
\right]
+
\epsilon_r,
\qquad
\epsilon_r>0,
\label{eq:warp_rate}
\end{equation}
where $g_{\boldsymbol{\phi}}(x)$ denotes the neural-network output,
$\boldsymbol{\phi}$ contains the trainable network parameters, and
$\epsilon_r$ is a small positive constant that prevents the rate from
vanishing.

The positive rate field is integrated to form a monotonic cumulative
component,
\begin{equation}
I(x)
=
\int_{x_{\min}}^{x}
r_{\boldsymbol{\phi}}(\xi)\,\mathrm{d}\xi ,
\label{eq:warp_integral}
\end{equation}
which is evaluated numerically on the ordered input samples using the
trapezoidal rule. The resulting values are standardized to zero mean
and unit variance,
\begin{equation}
\widetilde{I}(x_i)
=
\frac{
I(x_i)-\langle I\rangle
}{
{\rm std}(I)
}.
\label{eq:warp_integral_standardized}
\end{equation}
An intermediate warped coordinate is then constructed as
\begin{equation}
v(x_i)
=
\alpha x_i
+
\lambda_u \widetilde{I}(x_i),
\qquad
\alpha>0,
\quad
\lambda_u>0,
\label{eq:raw_warp}
\end{equation}
and is standardized once more to define the coordinate supplied to the
covariance kernel,
\begin{equation}
u(x_i)
=
\frac{
v(x_i)-\langle v\rangle
}{
{\rm std}(v)
}.
\label{eq:warped_coordinate}
\end{equation}

Because $r_{\boldsymbol{\phi}}(x)>0$, its cumulative integral is
monotonic in $x$. Together with the positivity constraints
$\alpha>0$ and $\lambda_u>0$, this ensures that $v(x)$ is monotonic.
The final affine standardization in
Equation~\eqref{eq:warped_coordinate} preserves this ordering.
Consequently, the learned transformation can modify the local spacing
of the time coordinate seen by the stationary base kernel without
reversing the temporal ordering of the observations. The linear term
provides a global monotonic baseline, whereas the integrated neural
component introduces smooth local departures from a uniform temporal
scale.

The transformation is deterministic conditional on the network
parameters and is optimized jointly with the GP covariance
hyperparameters. It therefore provides a learned input representation
for the covariance function rather than an additional latent Gaussian
process.

To quantify the local deformation of the learned coordinate, we define
\begin{equation}
\rho(x)
=
\frac{{\rm d}u}{{\rm d}x}.
\label{eq:deformation_rate}
\end{equation}
Here, $\rho(x)$ is the local stretching rate of the final warped
coordinate and should be distinguished from the auxiliary neural rate
$r_{\boldsymbol{\phi}}(x)$ in Equation~\eqref{eq:warp_rate}. If the
warp varies slowly over a local correlation scale, a characteristic
scale $\tau_u$ defined in the warped coordinate corresponds
approximately to the physical-time scale
\begin{equation}
\tau_{\rm eff}(t)
\simeq
\tau_u
\frac{s_t}{
\rho[x(t)]
},
\label{eq:local_physical_scale}
\end{equation}
where $s_t$ is the time-standardization factor introduced in
Equation~\eqref{eq:model_time}. The same conversion is applied to the
Matérn-$3/2$ correlation scale. Thus, larger values of $\rho$ imply a
shorter effective correlation scale in physical time, whereas smaller
values imply a longer one.

For post-fit diagnostics, the deformation rate is converted to the
physical-time coordinate according to
\begin{equation}
d_i
\equiv
\left.
\frac{{\rm d}u}{{\rm d}t}
\right|_{t_i}
=
\frac{1}{s_t}
\left.
\frac{{\rm d}u}{{\rm d}x}
\right|_{x_i}
=
\frac{\rho(x_i)}{s_t}.
\label{eq:warp_feature_rate}
\end{equation}
The sequence $\{d_i\}$ therefore provides a physical-time diagnostic
of the local temporal deformation inferred by the deep-kernel model.

\subsection{Optimization and regularization}
\label{sec:optimization}

For the nonstationary models, the GP covariance hyperparameters and the
parameters governing the corresponding nonstationary components are
optimized jointly. In the change-point models, these include the parameters
of the transition gate, whereas in the deep-kernel models the neural-network
parameters defining the learned time transformation are optimized together
with the base-kernel hyperparameters. The GP calculations are implemented
with \texttt{tinygp} \citep{tinygp2024} using the \texttt{JAX} framework
\citep{Bradbury2018}, and all trainable parameters are optimized using the
Adam algorithm \citep{KingmaBa2015}.

The optimization minimizes a regularized negative GP log-likelihood,
\begin{equation}
\mathcal{J}(\boldsymbol{\Theta})
=
-\ln \mathcal{L}(\boldsymbol{\Theta})
+
\mathcal{R}(\boldsymbol{\Theta}),
\label{eq:regularized_objective}
\end{equation}
where $\boldsymbol{\Theta}$ denotes the complete set of trainable parameters
and $\mathcal{R}$ contains weak regularization terms applied to the
nonstationary components. Because the objective is non-convex, each
nonstationary model configuration is optimized from three independent
random initializations. Model comparison is based on the unregularized
negative log-likelihood evaluated after optimization.

The regularization is intended to suppress poorly constrained or
numerically unstable nonstationary solutions without fixing the form of the
inferred evolution. For the deep-kernel models, the monotonic construction
described in Section~\ref{sec:deep_kernel_gp} is supplemented by weak
penalties that discourage local reversals, excessively rapid variations in
the deformation rate, and unnecessarily large nonlinear departures from a
near-uniform time coordinate. For the change-point models, the gate centre
and sharpness are bounded to physically admissible ranges, with additional
weak penalties discouraging transitions close to the boundaries of the
observed interval and excessively sharp or irregular gate profiles.

\subsection{Model assessment and residual diagnostics}
\label{sec:model_assessment}

After optimization, the unregularized negative log-likelihood is evaluated
for each completed run,
\begin{equation}
\mathrm{NLL}_r
=
-\ln p\!\left(\mathbf{y}\mid\boldsymbol{\Theta}^{*}_r\right),
\label{eq:nll}
\end{equation}
where $\boldsymbol{\Theta}^{*}_r$ denotes the optimized parameter vector
obtained from the $r$th random initialization. For each model configuration,
the mean NLL over the three runs is reported in Table~1 as the summary fit
statistic. The run with an NLL closest to this mean is adopted as the
representative fit for subsequent analysis.

At the observed times, we evaluate the conditional GP mean
$\widehat{y}_i$ and the conditional latent variance $V_i$. The total
variance used to normalize the residuals is
\begin{equation}
S_i^2
=
V_i+\sigma_i^2+j_i^2,
\end{equation}
and the standardized residuals are
\begin{equation}
e_i
=
\frac{y_i-\widehat{y}_i}{S_i}.
\label{eq:standardized_residuals}
\end{equation}
We examine the autocorrelation functions (ACFs) of both $e_i$ and
$e_i^2$ to diagnose remaining temporal correlation and residual conditional-variance structure, respectively \citep{McLeodLi1983}.
We additionally summarize these residual dependencies using Ljung--Box (LB) portmanteau tests over the adopted lag range
\citep{LjungBox1978}. Because the LB statistic combines information across multiple lags, its interpretation is combined with the individual ACF coefficients in the assessment of residual model adequacy.

As a restricted conditional check of the covariance structure at the gate-defined boundary, we additionally perform a separately refitted fixed-split Bayesian evidence analysis. The split boundary is inherited from the representative gate fit and is held fixed throughout this comparison; it is not re-optimized within the fixed-split analysis. The resulting evidence comparison is therefore conditional on the gate-defined boundary and does not constitute an independent search for a change point.

Stationary DRW and M32 covariance models are fitted independently on the two sides of the fixed boundary. With these two candidate covariance families, the fixed-split analysis therefore considers four segment-wise assignments: DRW$\rightarrow$DRW, DRW$\rightarrow$M32, M32$\rightarrow$DRW, and M32$\rightarrow$M32. The same stationary candidates are also fitted to the full exposure to provide the reference evidence. This construction tests whether allowing the covariance description to vary between the two temporal segments is supported relative to a stationary full-exposure description.

For all stationary fits used in the fixed-split comparison, the GP
amplitude and characteristic correlation scale are sampled in
logarithmic coordinates. Uniform priors are adopted for
$\ln\sigma$ and for the logarithm of the corresponding correlation-scale
parameter, equivalent to log-uniform priors on the physical parameters.
We use
\begin{equation}
\sigma\in[0.1,200]~{\rm counts~s^{-1}},
\qquad
\tau,\ell\in[0.01,150]~{\rm s},
\end{equation}
with identical prior ranges for the two sub-intervals and the
full-exposure reference fits. The additional white-noise jitter term
$j$ is fixed to zero in these nested-sampling fits and is not included
as a sampled parameter. The GP mean is fixed to the full-exposure
sample mean for all compared fits, providing a common mean prescription
so that the evidence comparison primarily reflects differences in the
covariance structure.

The Bayesian evidences used in the fixed-split analysis are computed
with \texttt{dynesty}, a dynamic nested-sampling algorithm designed for
Bayesian posterior exploration and marginal-likelihood estimation
\citep{Speagle2020}. For each stationary fit, \texttt{dynesty} provides
an estimate of the Bayesian evidence $Z$, which is used here to compare
the segmented and full-exposure covariance descriptions.

For a specified pair of stationary covariance models,
$K_{\rm pre}$ and $K_{\rm post}$, fitted on the two sides of the fixed
boundary, we define the fixed-split evidence gain as
\begin{equation}
\Delta\ln Z_{\rm split}
=
\ln Z_{\rm pre}(K_{\rm pre})
+
\ln Z_{\rm post}(K_{\rm post})
-
\ln Z_{\rm full},
\label{eq:fixed_split_evidence}
\end{equation}
where $\ln Z_{\rm full}$ denotes the Bayesian evidence of the preferred
stationary model fitted to the full exposure. A positive
$\Delta\ln Z_{\rm split}$ indicates that, conditional on the adopted
fixed boundary, the segmented covariance description is favoured over
the stationary full-exposure reference.

The uncertainty reported for $\Delta\ln Z_{\rm split}$ is obtained by
propagating the evidence-sampling uncertainties returned by
\texttt{dynesty} for the pre-transition, post-transition, and
full-exposure fits. These nested-sampling runs are treated as
independent in the uncertainty propagation.

\section{Results}
\label{sec:results}

\subsection{Overall performance of the nonstationary models}
\label{sec:overall_performance}

We first apply the four nonstationary GP models to the fiducial
10--300~keV full-band analysis with 0.10~s time bins. The models comprise two change-point
configurations, corresponding to DRW-to-Matérn-$3/2$ and
Matérn-$3/2$-to-DRW covariance transitions, together with two
deep-kernel models constructed from DRW and Matérn-$3/2$ base
covariances. These models provide complementary descriptions of
nonstationarity, either through an explicit transition between two
covariance regimes or through a continuous deformation of the time
coordinate.

\begin{figure}[htbp]
\centering

\includegraphics[
width=0.88\textwidth
]{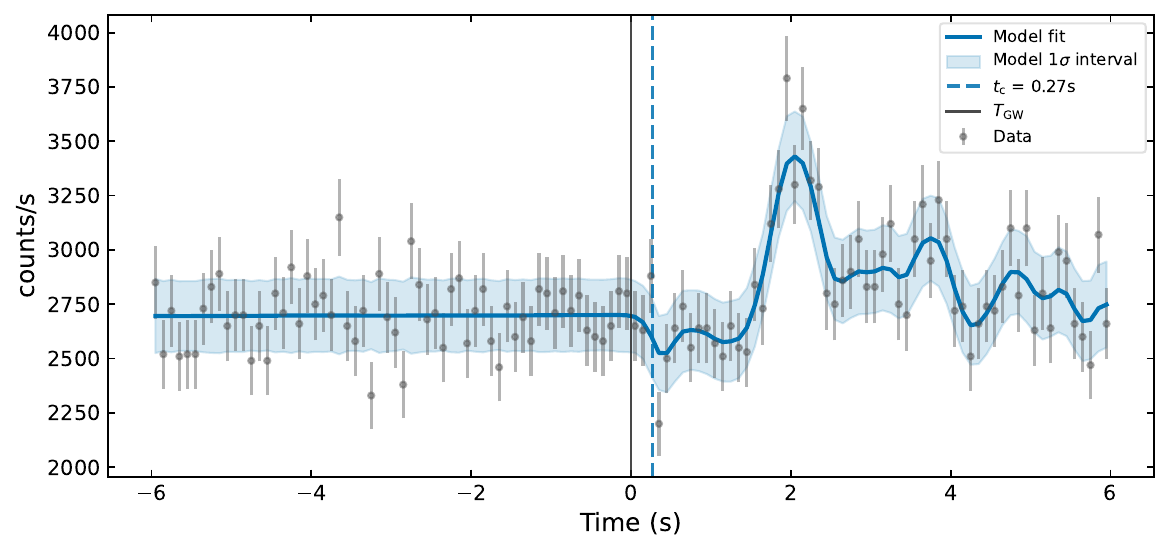}

\vspace{0.3em}

\includegraphics[
width=0.88\textwidth
]{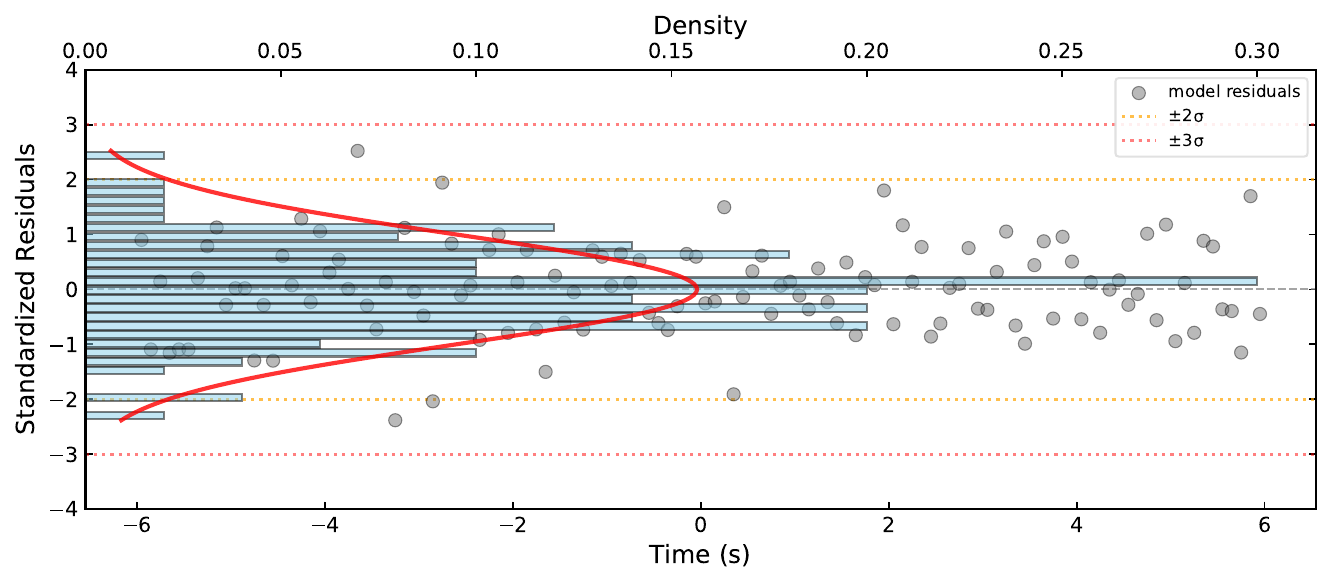}

\vspace{0.5em}

\begin{minipage}[b]{0.48\textwidth}
\centering
\includegraphics[
width=\linewidth
]{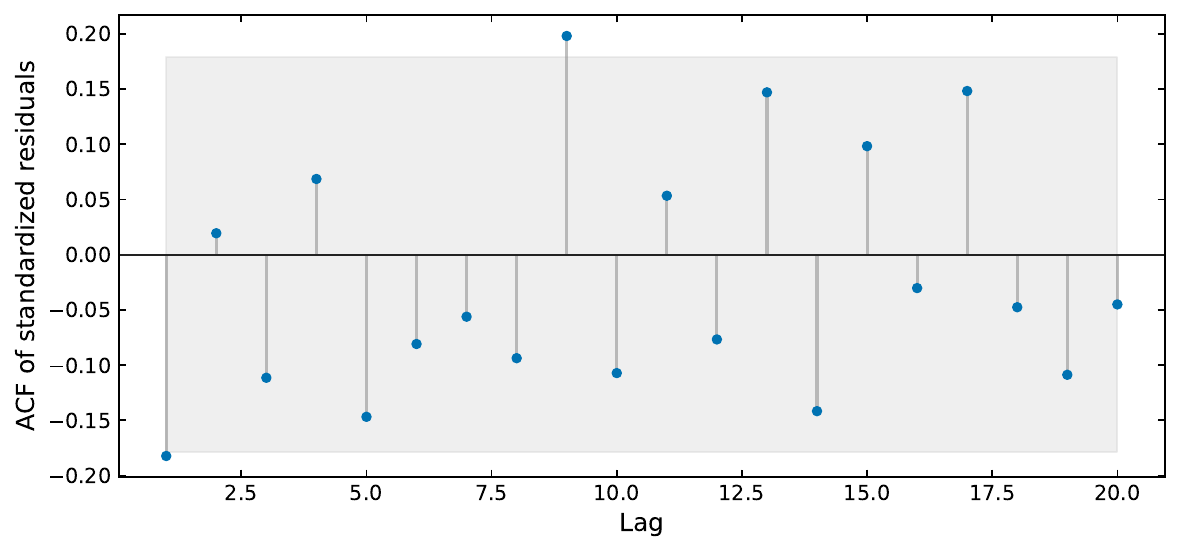}
\end{minipage}
\hfill
\begin{minipage}[b]{0.48\textwidth}
\centering
\includegraphics[
width=\linewidth
]{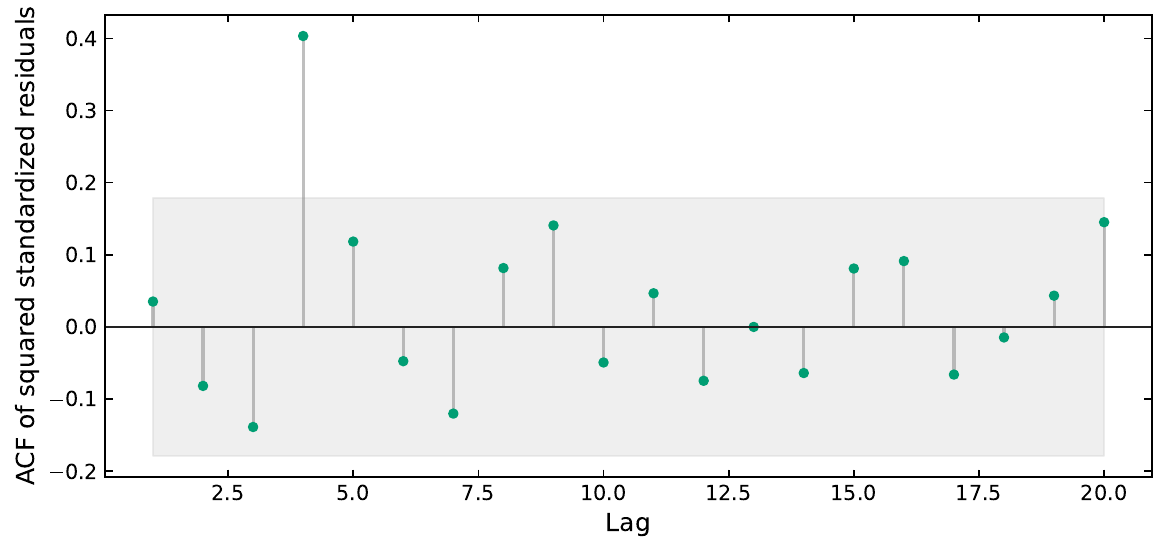}
\end{minipage}

\caption{
Representative fit and residual diagnostics for the
DRW$\rightarrow$\matern{} change-point model.
The top panel shows the GP fit and its $1\sigma$ model interval; the
solid and dashed vertical lines mark $T_{\rm GW}=0$ and the fitted gate
centre, respectively. The middle panel shows the standardized residuals
and their marginal distribution. The lower panels show the
autocorrelation functions of the standardized residuals and their
squares, with shaded regions indicating the approximate
zero-correlation intervals.
}
\label{fig:gate_fit_acf_stability}
\end{figure}

Figure~\ref{fig:gate_fit_acf_stability} shows the
DRW-to-Matérn-$3/2$ change-point model as a representative example.
The fitted GP reproduces the main temporal structure of the prompt
emission across the full exposure, while the standardized residuals
remain distributed around zero without broad systematic departures.
The residual autocorrelation functions further show that most of the
temporal correlation present in the original light curve has been
absorbed by the fitted covariance model. Most individual ACF
coefficients lie within the approximate zero-correlation intervals,
although some localized residual structure remains, most noticeably
in the squared-residual ACF around lag~4.

The corresponding fit statistics and residual diagnostics for all
four nonstationary models are summarized in
Table~\ref{tab:modelsummary}. Overall, each model provides a viable
description of the dominant temporal variability in the baseline
light curve. The change-point models achieve somewhat better
in-sample likelihoods, whereas the deep-kernel models generally leave
weaker correlations in the squared residuals. The residual diagnostics
therefore indicate modest differences in how the two classes of models
capture the remaining short-timescale structure, rather than a
qualitative failure of any individual model.

The likelihood and residual diagnostics show that both
forms of nonstationary GP model capture the principal temporal
correlations of the prompt-emission light curve. The following
sections therefore focus on the nonstationary features inferred by
these models, in particular the timing and stability of the
change-point and deep-kernel signatures.

\input{model_summary_0p1_10_300.tex}

\subsection{Covariance transition close to the merger epoch}
\label{sec:change_point_results}

Both change-point configurations identify an early covariance
transition in the fiducial 10--300~keV full-band analysis with
0.10~s time bins.
For temporal comparison with the prompt emission, we adopt the
reported GW--gamma-ray delay as the gamma-ray onset reference,
$\tgamma=1.74~{\rm s}$ after $T_{\rm GW}$
\citep{Abbott2017PRL}.

\begin{figure}[htbp]
\centering

\includegraphics[
width=0.78\textwidth
]{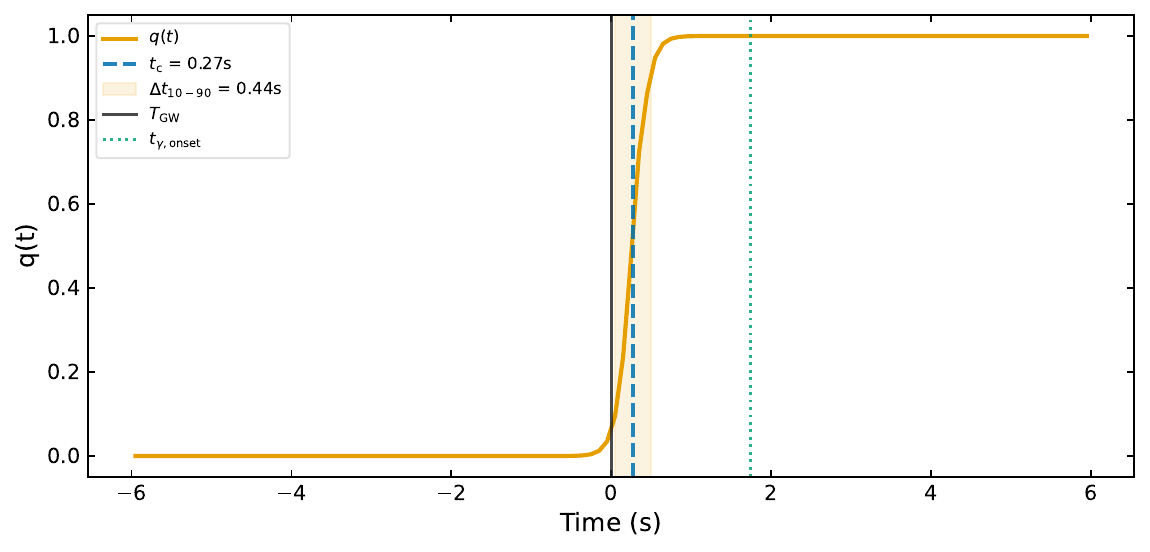}

\caption{
Sampled gate function for the representative
DRW$\rightarrow$\matern{} change-point model. The dashed line marks the
parametric gate centre $t_{\rm c}$, and the shaded interval extends
from the numerically interpolated $q(t)=0.1$ crossing to the
$q(t)=0.9$ crossing. The solid black and dotted green lines mark $T_{\rm GW}$
and $\tgamma$, respectively.
}
\label{fig:diag_gate_drw_to_m32}
\end{figure}

Figure~\ref{fig:diag_gate_drw_to_m32} shows the sampled gate function
$q(t)$ for the representative DRW$\rightarrow$\matern{}
configuration. The gate evolves smoothly from the pre-transition to
the post-transition covariance regime close to the merger epoch. The
numerically determined $q(t)=0.1$--$0.9$ interval characterizes the
effective extent of the transition and shows that the main covariance
evolution occurs well before the adopted gamma-ray onset reference.

The gate timing quantities for both change-point configurations are
summarized in Table~\ref{tab:gate}. Despite the opposite ordering of
their covariance components, the DRW$\rightarrow$\matern{} and
\matern{}$\rightarrow$DRW models recover closely similar transition
epochs and extents. In both cases, the inferred covariance evolution occurs within the
early post-merger interval and is largely completed before the adopted
gamma-ray onset reference.

\input{gate_timing_0p1_10_300.tex}

We next perform the restricted fixed-split analysis at
$t_{\rm tr}=0.27~{\rm s}$, corresponding to the representative
DRW$\rightarrow$\matern{} gate centre. The split boundary is held
fixed throughout this comparison. Stationary DRW and \matern{}
covariance models are then fitted independently to the data before
and after the boundary, together with stationary fits to the full
exposure.

\begin{figure}[htbp]
\centering

\includegraphics[
width=0.90\textwidth
]{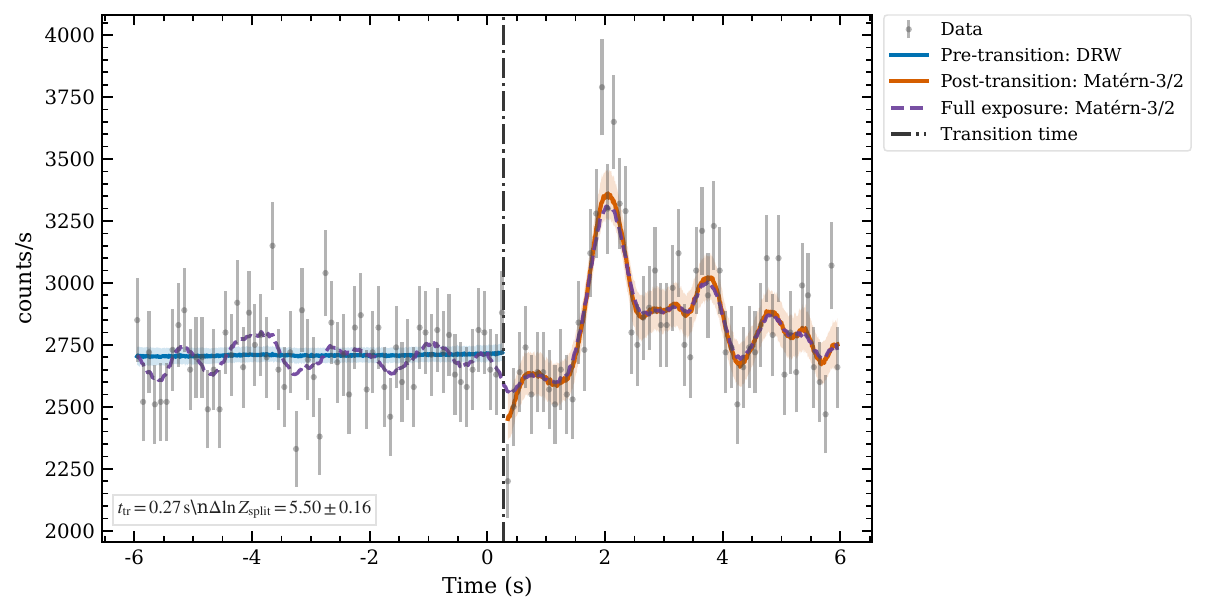}

\caption{
Representative fixed-split diagnostic at the gate-defined boundary
$t_{\rm tr}=0.27~{\rm s}$. The pre-transition and post-transition
curves show the independently refitted stationary DRW and
Mat\'ern-$3/2$ GPs for the representative
DRW$\rightarrow$Mat\'ern-$3/2$ assignment. The stationary
full-exposure Mat\'ern-$3/2$ fit is shown for comparison, and the
vertical dot-dashed line marks the fixed boundary.
}
\label{fig:fixed_split_diagnostic}
\end{figure}

Figure~\ref{fig:fixed_split_diagnostic} shows the
DRW$\rightarrow$\matern{} assignment as the representative
fixed-split configuration. This choice directly corresponds to the
covariance ordering of the gate model from which the adopted boundary
is obtained. The figure compares the independently refitted
stationary models on the two temporal segments with the stationary
full-exposure \matern{} reference.

The Bayesian evidences of the stationary models fitted to the two
segments and to the full exposure are summarized in
Table~\ref{tab:fixed_split_evidence}. Among the full-exposure
stationary reference models, the \matern{} model has the higher
Bayesian evidence and is therefore used as $\ln Z_{\rm full}$ in
Equation~\eqref{eq:fixed_split_evidence}.

\input{fixed_split_evidence.tex}

All four segment-wise assignments defined in Section~\ref{sec:model_assessment} yield positive
$\Delta\ln Z_{\rm split}$ at this fixed boundary. The largest evidence gain is obtained for the
Mat\'ern-$3/2\rightarrow$Mat\'ern-$3/2$ assignment, with
$\Delta\ln Z_{\rm split}=5.72\pm0.16$. The fact that the largest gain
is obtained without changing the covariance family shows that the
conditional fixed-split evidence does not require a transition between
different covariance families: allowing the covariance parameters to
vary between the two temporal segments is itself sufficient to improve
the description relative to the full-exposure stationary reference.

For direct correspondence with the gate model from which the adopted
boundary is obtained, we use the
DRW$\rightarrow$Mat\'ern-$3/2$ assignment as the representative
fixed-split configuration. This assignment gives
$\Delta\ln Z_{\rm split}=5.50\pm0.16$, close to the maximum obtained
for the Mat\'ern-$3/2\rightarrow$Mat\'ern-$3/2$ case. Conditional on
the gate-defined boundary, the fixed-split comparison therefore
supports segment-specific covariance structure, but does not uniquely
identify a transition between different covariance families.

\subsection{Localized time deformation near the $\gamma$-ray onset}
\label{sec:deep_kernel_results}

We characterize the deep-kernel timing structure using the physical-time
deformation rate $\dd u/\dd t$ defined in
Section~\ref{sec:deep_kernel_gp}. Because its absolute amplitude is
model dependent, the location and descriptive extent of the dominant
localized feature are used as the primary timing diagnostics. The dominant
feature is identified as the largest local maximum of the sampled
$\dd u/\dd t$ curve over the analysed interval, with the surrounding
contiguous rising and falling portions defining $t_{\rm start}$ and
$t_{\rm end}$, respectively.

\begin{figure}[htbp]
\centering

\includegraphics[
width=0.82\textwidth
]{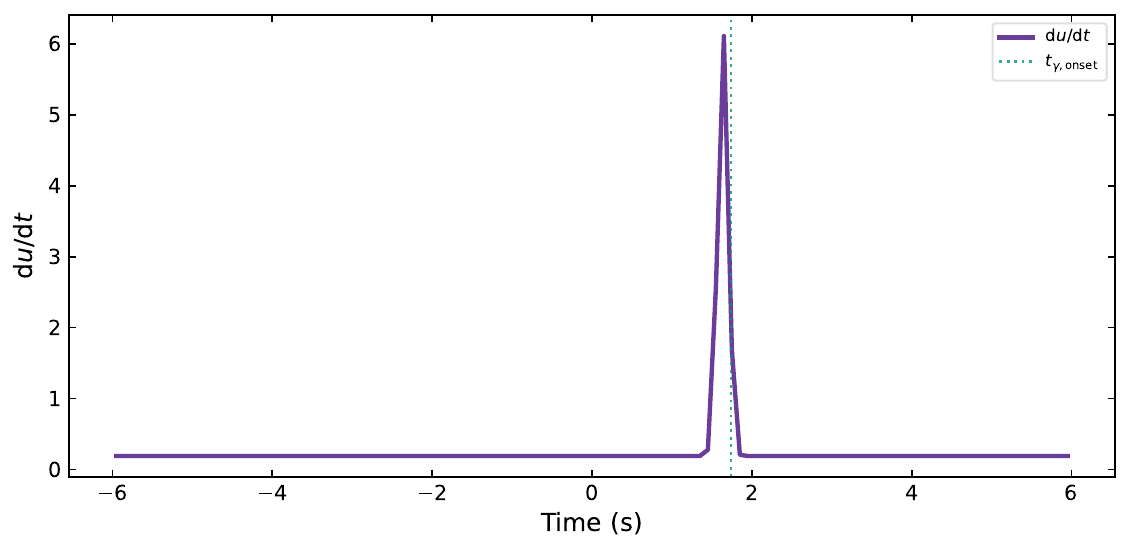}

\vspace{0.5em}

\includegraphics[
width=0.82\textwidth
]{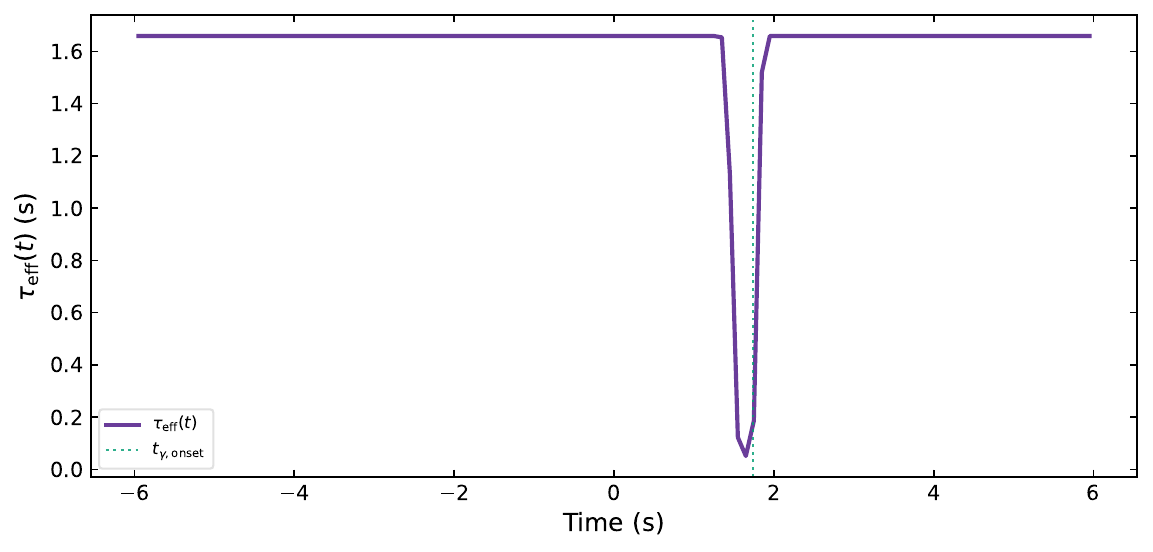}

\caption{
Deep-kernel timing diagnostics for the representative DRW warp model fitted to the 10--300 keV light curve with 0.10 s time bins. The upper panel shows
the learned physical-time deformation rate $\dd u/\dd t$, and the lower
panel shows the corresponding DRW effective correlation time
$\tau_{\rm eff}(t)$. The dotted vertical line marks $\tgamma$. The localized maximum of $du/dt$ corresponds to a temporary reduction
in $\tau_{\rm eff}(t)$.
}
\label{fig:diag_warp_drw}
\end{figure}

Figure~\ref{fig:diag_warp_drw} shows the representative DRW
deep-kernel diagnostics. The learned deformation rate exhibits a localized
enhancement near the adopted gamma-ray onset reference. The corresponding
effective correlation time, $\tau_{\rm eff}(t)$, decreases over the same
interval, illustrating the local change in temporal correlation scale
associated with the learned time deformation. The maximum of
$\dd u/\dd t$ provides the representative warp-based timing marker.

The timing quantities and fitted warp parameters for both deep-kernel
configurations are summarized in Table~\ref{tab:warp}. Both the DRW and
\matern{} base-kernel models recover a localized deformation feature near
the gamma-ray onset, with closely spaced peak locations despite their
different underlying covariance families. In both cases, the peak offset
from $\tgamma$ is within the 0.10~s temporal sampling resolution.

Neither the model optimization nor the post-fit feature-selection procedure
uses $t_{\gamma,\rm onset}$ as an input. The proximity between the recovered
deformation peaks and the gamma-ray onset is therefore treated as a post-fit
empirical correspondence. The reported
$t_{\rm start}$--$t_{\rm peak}$--$t_{\rm end}$ ranges characterize the
descriptive extent of the localized deformation and are not statistical
confidence or credible intervals.

\input{warp_timing_0p1_10_300.tex}

\subsection{Stability under energy selection and time binning}
\label{sec:stability_results}

\begin{figure}[htbp]
\centering

\includegraphics[
width=0.82\textwidth
]{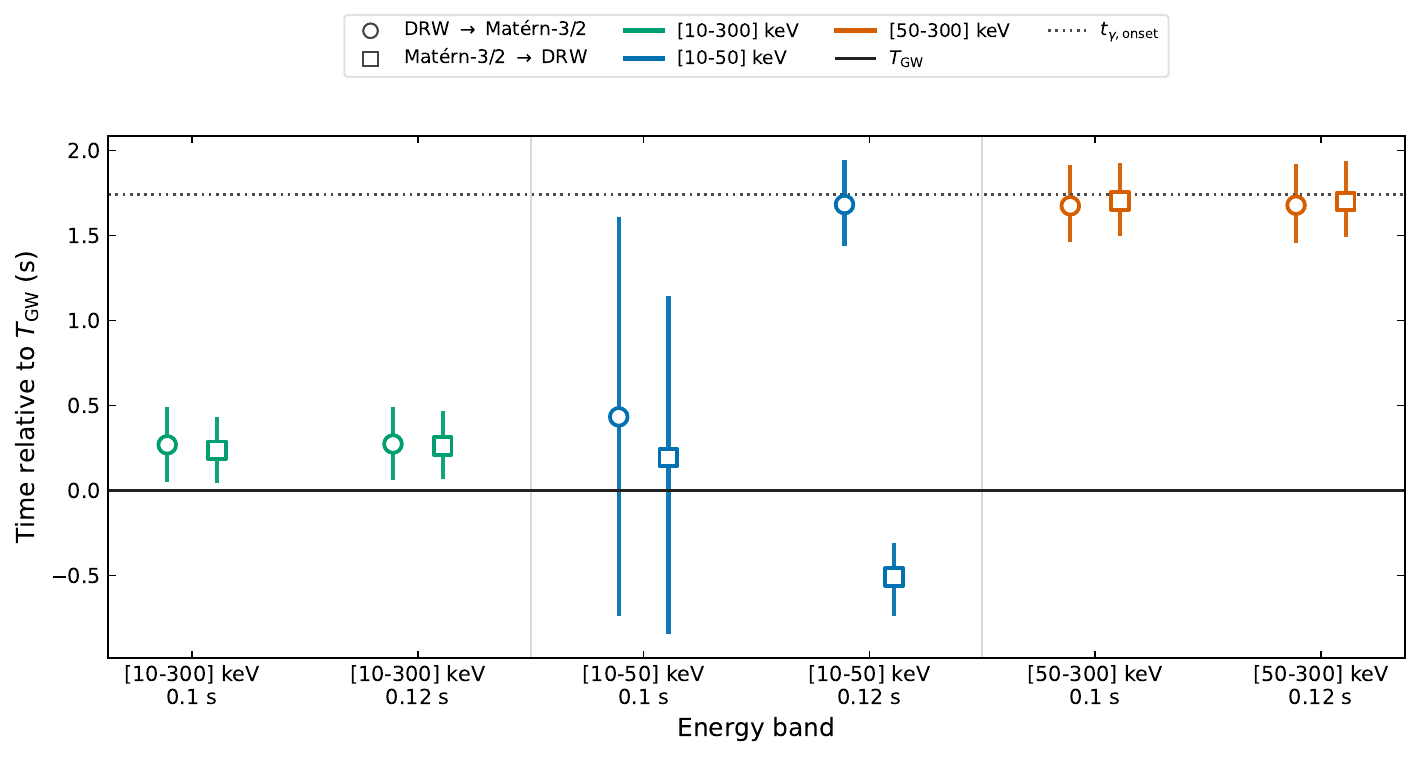}

\vspace{0.5em}

\includegraphics[
width=0.82\textwidth
]{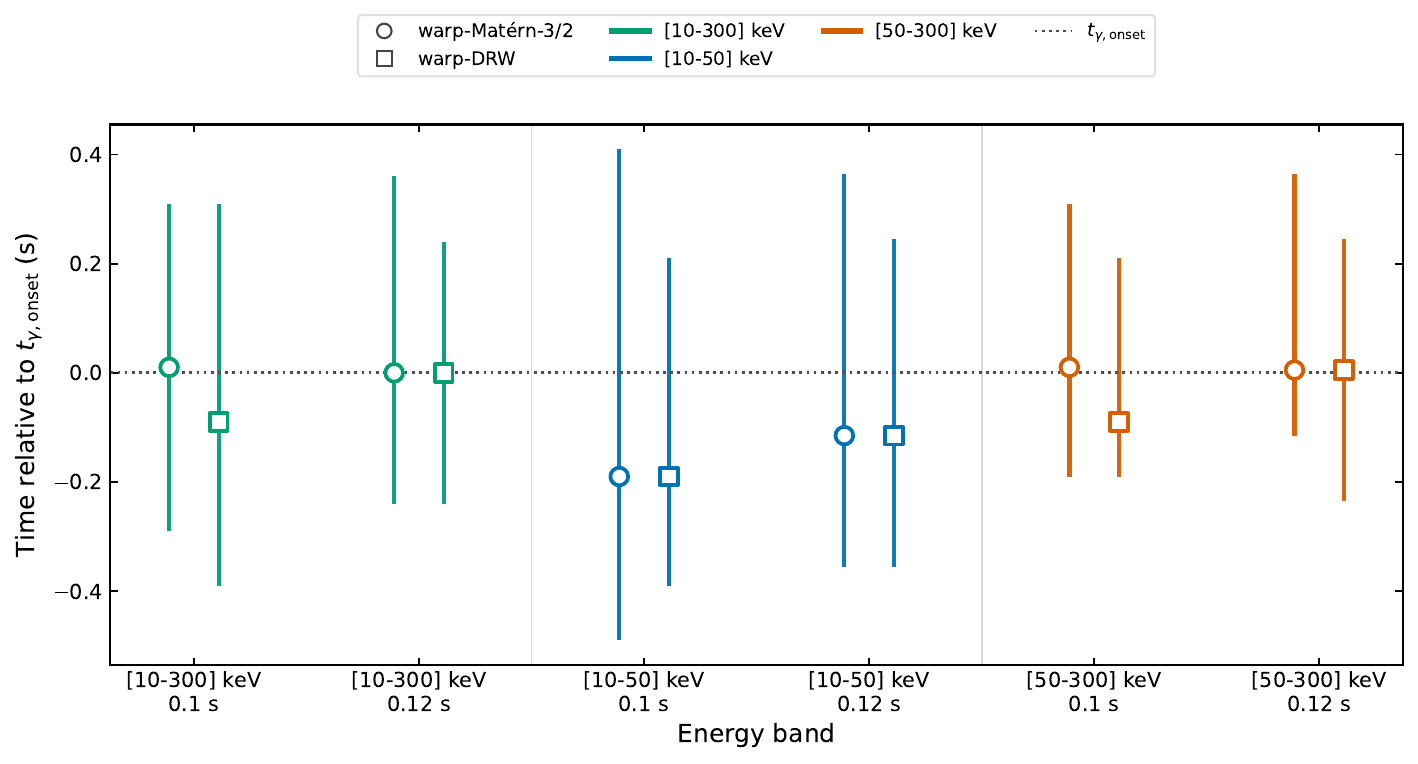}

\caption{
Stability of the fitted timing diagnostics under changes in energy selection and time binning. Upper: gate timing relative to
$T_{\rm GW}$. Circular and square markers show the parametric gate centres $t_{\rm c}$ for the
DRW$\rightarrow$\matern{} and
\matern{}$\rightarrow$DRW configurations, respectively, with vertical segments spanning the numerically determined
$q(t)=0.1$--$0.9$ transition extents. Lower: warp timing relative to the adopted gamma-ray onset, showing
$t_{\rm peak}-t_{\gamma,\rm onset}$ and the corresponding start--end deformation extents. Colours identify the energy bands consistently
between the two panels. All vertical extents are descriptive fitted
feature ranges rather than statistical confidence or credible intervals.
}
\label{fig:cross_band_timing}
\end{figure}

We repeat the change-point and deep-kernel timing analyses for the 10--50 and 50--300~keV sub-bands and
for an alternative time resolution of $\Delta t=0.12$~s. These calculations provide stability tests under changes in energy selection and temporal binning. The 10--300~keV full-band analysis with 0.10~s time bins remains
the fiducial configuration.
Figure~\ref{fig:cross_band_timing} provides an overview of the recovered
gate and warp timings across the tested analysis configurations.

The full-band timing results are stable to the modest change in
temporal binning. At $\Delta t=0.12$~s, the
DRW$\rightarrow$\matern{} and \matern{}$\rightarrow$DRW gate
configurations give parametric centres of 0.274 and 0.262~s after
$T_{\rm GW}$, with numerical 10--90\% transition intervals of
0.060--0.491 and 0.066--0.469~s, respectively. These values are close
to the corresponding 0.10~s full-band centres of 0.269 and 0.237~s.
The full-band warp timing is similarly stable: at 0.12~s resolution,
both deep-kernel models recover deformation peaks at 1.74~s after
$T_{\rm GW}$, compared with 1.65 and 1.75~s for the fiducial
0.10~s DRW and \matern{} fits. The corresponding 0.12~s start--end
feature extents are 1.50--1.98 and 1.50--2.10~s.

A substantially larger variation is found when the data are divided into energy sub-bands, particularly for the gate diagnostic. In the 10--50~keV band, the inferred gate locations vary strongly with both model configuration and time binning: the 0.10~s fits have broad transition intervals, while the two 0.12~s configurations place one gate centre before $T_{\rm GW}$ and the other close to the observed gamma-ray onset. In the 50--300~keV band, by contrast, the two gate configurations place their centres near the onset reference. The sub-band gate results are therefore substantially less consistent than the corresponding full-band rebinning results.

Two effects may contribute to this behaviour. First, dividing the full-band data into narrower energy intervals changes the signal-to-noise ratio. The
reduced statistical information available to each sub-band fit may
alter which covariance structure is most strongly constrained and
consequently shift the inferred gate location. Second, the observed
differences may contain a genuine energy-dependent component if the
temporal variability or relative contribution of emission structures
changes with photon energy. The present analysis does not distinguish
between these statistical and physical contributions.

The warp feature shows a smaller variation under the same tests.
Across the tested sub-bands and time resolutions, the deformation
peaks remain within approximately 1.55--1.75~s after $T_{\rm GW}$.
The soft-band solutions tend to occur somewhat earlier than
$\tgamma$, whereas the full-band and hard-band solutions
remain more closely concentrated around the onset reference. The warp timing shows a smaller cross-band variation than the gate timing under the tested energy selections. The remaining shifts nevertheless show that the recovered warp feature is not strictly invariant with energy selection.

\section{Discussion}
\label{sec:discussion}

\subsection{Two phenomenological timing signatures}
\label{sec:timing_interpretation}

The fiducial analysis reveals two distinct phenomenological timing
signatures in the stochastic structure of the GRB~170817A light curve.
The change-point models recover an early covariance transition within the
first few tenths of a second after $T_{\rm GW}$, while the deep-kernel
models identify a later maximum in the learned time-deformation rate near
the observed gamma-ray onset. These features arise from different
nonstationary constructions and should therefore be regarded as
complementary diagnostics rather than estimates of the same transition
time.

The early feature is recovered for both DRW$\rightarrow$\matern{} and
\matern{}$\rightarrow$DRW orderings, indicating that its timing is not
set by a particular covariance assignment. The deep-kernel feature is
also recovered with both DRW and \matern{} base kernels, with deformation
maxima at 1.65 and 1.75~s after $T_{\rm GW}$, respectively, close to the
adopted gamma-ray onset at $t_{\gamma,\rm onset}=1.74$~s. This
correspondence emerges only after model fitting and post-fit feature
extraction.

Neither feature can be uniquely associated with a specific radiation
mechanism or dynamical stage. Instead they provide phenomenological timing constraints on the evolution of the stochastic covariance structure that can be compared with physical scenarios for the prompt-emission epoch \citep{Gottlieb2018,Duffell2018,Geng2019}.

\subsection{Fit diagnostics and robustness}
\label{sec:robustness_discussion}

All four nonstationary models reproduce the dominant structure of the fiducial light curve. The change-point models provide the smaller mean in-sample negative log-likelihoods across the completed random initializations, whereas the deep-kernel models show more favourable squared-residual diagnostics. The fixed-split comparison at 0.27 s provides an additional conditional check of the covariance structure at the gate-defined boundary. All four segment-wise assignments yield positive evidence gains relative to the full-exposure \matern{} reference, with $\Delta\ln Z_{\rm split}$ spanning approximately $4.54$--$5.72$. The largest gain is obtained for the \matern{}$\rightarrow$\matern{} assignment ($5.72\pm0.16$), while the gate-matched DRW$\rightarrow$\matern{} assignment gives $5.50\pm0.16$. Conditional on the adopted gate-defined boundary, the fixed-split result therefore favours segment-specific covariance structure without uniquely identifying a change of covariance family.

The residual diagnostics nevertheless show that some localized time-dependent variance remains unresolved, particularly in the squared residuals of the change-point fits. This residual structure
indicates that the fitted covariance evolution does not capture every component of the prompt-emission variability, while the principal gate
and warp timing features remain localized within their respective model constructions.

The stability analysis shows different responses to temporal binning and energy selection. In the full 10--300~keV band, both timing
diagnostics remain stable when the bin width is changed from 0.10 to 0.12~s. Energy subdivision produces substantially larger variation in
the gate locations, while the warp peaks remain within the narrower range of approximately 1.55--1.75~s after $T_{\rm GW}$. The larger cross-band variation of the gate timing may reflect both the changed photon statistics and signal-to-noise ratio of the sub-band light
curves and genuine energy-dependent temporal structure; the present analysis does not separate these contributions. The full-band gate result is therefore retained as the primary early timing measurement, while the later warp feature shows greater stability across the tested analysis configurations.

\subsection{Limitations}
\label{sec:limitations}

The present analysis adopts a Gaussian likelihood for the
background-subtracted, binned count-rate light curve. The Gaussian approximation may become less accurate in intervals with limited photon counts. A joint count-level treatment of source and background events could provide a more complete statistical description.

A more extensive calibration using stationary and nonstationary simulations is required to quantify false-positive rates, timing bias, and parameter-recovery accuracy. The correspondence of the recovered timing features with $T_{\rm GW}$ and $t_{\gamma,\rm onset}$ is therefore treated as empirical within the present data set rather than as a calibrated statistical significance. In particular, the early gate-defined boundary precedes the observed gamma-ray onset, and the present analysis cannot establish whether this feature originates from intrinsic source variability, the measurement and background process, or model dependence. In addition, the change-point and deep-kernel constructions are analysed separately in this work; future joint modelling could examine whether discrete covariance evolution and continuous time deformation can be represented simultaneously, although such an extension would require additional validation.

\section{Conclusions}
\label{sec:conclusions}

We have used change-point and deep-kernel GP models in the fiducial
10--300~keV full-band analysis with 0.10~s time bins to characterize
time-dependent stochastic variability in GRB~170817A. The two nonstationary constructions describe complementary forms of covariance
evolution. The change-point models represent a localized transition
between covariance descriptions, whereas the deep-kernel models
describe continuous changes in the effective temporal correlation
structure through a learned time deformation.

The change-point models place an early covariance transition at 0.269
and 0.237 s after $T_{\rm GW}$, with the corresponding 10--90\%
transition extents largely confined to the first $\sim0.5$ s.
Conditional on the representative gate-defined boundary at
$t_{\rm tr}=0.27$ s, all four fixed-split covariance assignments give
positive evidence gains over the stationary full-exposure reference.
The largest gain is obtained for the
Mat\'ern-3/2$\rightarrow$Mat\'ern-3/2 assignment. The fixed-split
analysis therefore supports segment-specific covariance structure
without requiring a uniquely identified change of covariance family.

The deep-kernel models recover a later localized time-deformation
feature with peaks at 1.65 and 1.75 s after $T_{\rm GW}$, close to the
adopted gamma-ray onset at
$t_{\gamma,\rm onset}=1.74$ s. The corresponding deformation represents
a localized change in the effective temporal correlation scale. The
full-band gate and warp timings remain stable when the bin width is
changed from 0.10 to 0.12 s. Energy subdivision produces substantially
larger shifts in the gate locations, whereas the warp peaks remain
confined to a narrower late-time range.

Several aspects of the validation motivate a cautious interpretation.
Localized residual dependence remains, particularly in the squared
residuals of the change-point fits, and the inferred gate locations are
less stable under energy subdivision. The fixed-split evidence is
conditional on a boundary obtained from the gate model, and the
recovered timing features have not yet been calibrated using large
ensembles of stationary and nonstationary simulations. In particular,
because the early gate-defined feature precedes the observed gamma-ray
onset, the present analysis does not establish that it represents a
physical source-state transition.

We therefore regard the early gate boundary and the later warp feature
as model-dependent phenomenological stochastic timing diagnostics
rather than calibrated physical transition times. Their recovery with
two complementary nonstationary constructions demonstrates the
potential of nonstationary GP modelling for identifying temporal
structure that is not captured by a globally time-invariant covariance
description. Further count-level modelling and simulation-based
calibration will be required to determine the statistical robustness
and physical origin of these features.

\section*{Acknowledgments}
 D.Y. acknowledges support from the National Natural Science Foundation of China (grant No.~12393852) and the Yunnan Provincial Science and Technology Department Foundation (grant No.~202601AT070175). We acknowledge the public \textit{Fermi}/GBM archive and the Laser Interferometer Gravitational-Wave Observatory (LIGO)--Virgo Collaboration for making GW170817-related data products publicly available.

\bibliographystyle{elsarticle-harv}
\bibliography{refs}

\end{document}

%% file: model_summary_0p1_10_300.tex
\begin{table}[htbp]
\centering
\caption{
Summary diagnostics for the four nonstationary models. Reported NLL values are averaged over three completed random initializations and rounded to the nearest integer. The
Ljung--Box (LB) $p$-values are calculated for the representative run whose
NLL is closest to the corresponding three-run mean \citep{LjungBox1978}. 
}
\label{tab:modelsummary}

\scriptsize
\renewcommand{\arraystretch}{1.10}

\begin{tabular*}{\textwidth}{
@{\extracolsep{\fill}}
llccc
@{}
}
\toprule
Class & Scenario & NLL & LB residual $p$ & LB squared-residual $p$ \\
\midrule
warp & DRW warp
& 803
& 0.086 & 0.440 \\

warp & \matern{} warp
& 804
& 0.035 & 0.387 \\

gate & DRW$\rightarrow$\matern{} gate
& 797
& 0.044 & 0.006 \\

gate & \matern{}$\rightarrow$DRW gate
& 798
& 0.069 & $3.43\times10^{-4}$ \\
\bottomrule 
\end{tabular*}

\end{table}

%% file: gate_timing_0p1_10_300.tex
\begin{table}[htbp]
\centering
\caption{
Gate timing quantities for the two change-point configurations. Times are seconds relative to $T_{\rm GW}$. The gate
centre $t_{\rm c}$ is the parametric centre of the analytic
sigmoid component. The boundaries $t_{10}$ and $t_{90}$ are
determined by linear interpolation of the sampled fitted gate
curve at $q(t)=0.1$ and $q(t)=0.9$, respectively, and
$\Delta t_{10-90}=t_{90}-t_{10}$. No analytic width derived from
the fitted sharpness $s$ is used.
}
\label{tab:gate}

\scriptsize
\renewcommand{\arraystretch}{1.10}

\begin{tabular*}{\textwidth}{
@{\extracolsep{\fill}}
lccccc
@{}
}
\toprule
Scenario &
$t_{\rm c}$ &
$t_{10}$ &
$t_{90}$ &
$\Delta t_{10-90}$ &
$s$ \\
\midrule
DRW$\rightarrow$\matern{} gate
& 0.269 & 0.053 & 0.494 & 0.441 & 37.10 \\

\matern{}$\rightarrow$DRW gate
& 0.237 & 0.043 & 0.435 & 0.393 & 37.82 \\
\bottomrule
\end{tabular*}

\end{table}

%% file: fixed_split_evidence.tex
\begin{table}[htbp]
\centering
\caption{
Bayesian log evidences of the stationary covariance models fitted to the intervals defined by the fixed boundary $t_{\rm tr}=0.27~{\rm s}$ and to the full exposure. DRW and M32 denote the damped random walk and Matérn-$3/2$ covariance models, respectively.
}
\label{tab:fixed_split_evidence}

\scriptsize
\renewcommand{\arraystretch}{1.10}

\begin{tabular*}{\textwidth}{
@{\extracolsep{\fill}}
lcc
@{}
}
\toprule
Interval & $\ln Z_{\rm DRW}$ & $\ln Z_{\rm M32}$ \\
\midrule
Pre-transition  & $-409\pm0.060$ & $-409\pm0.061$ \\
Post-transition & $-397\pm0.103$ & $-396\pm0.107$ \\
Full exposure   & $-812\pm0.099$ & $-811\pm0.101$ \\
\bottomrule
\end{tabular*}

\end{table}

%% file: warp_timing_0p1_10_300.tex
\begin{table}[htbp]
\centering
\caption{
Warp timing quantities and fitted parameters for the full-band
10--300~keV light curve with 0.10 s time bins. The times
$t_{\rm start}$, $t_{\rm peak}$, and $t_{\rm end}$ are given in seconds
relative to $T_{\rm GW}$, and the onset offset is
$t_{\rm peak}-t_{\gamma,\rm onset}$ with
$t_{\gamma,\rm onset}=1.74~{\rm s}$. The reported start--end ranges
describe the extent of the selected deformation feature and are not
statistical uncertainty intervals.
}
\label{tab:warp}

\scriptsize
\setlength{\tabcolsep}{4pt}
\renewcommand{\arraystretch}{1.10}

\begin{tabular*}{\textwidth}{
@{\extracolsep{\fill}}
lccccccccc
@{}
}
\toprule
Scenario &
$t_{\rm start}$ &
$t_{\rm peak}$ &
$t_{\rm end}$ &
$t_{\rm peak}-\tgamma$ &
$\Delta t_{-}$ &
$\Delta t_{+}$ &
$(\dd u/\dd t)_{\rm peak}$ &
$\alpha$ &
$\lambda_u$ \\
\midrule
DRW warp
& 1.35 & 1.65 & 2.05 & -0.09
& 0.30 & 0.40 & 6.113 & 0.649 & 0.475 \\

\matern{} warp
& 1.45 & 1.75 & 2.05 & 0.01
& 0.30 & 0.30 & 2.727 & 1.308 & 0.163 \\
\bottomrule
\end{tabular*}

\parbox{\textwidth}{
\vspace{2pt}
\footnotesize
\textit{Note.}
Here $\Delta t_{-}=t_{\rm peak}-t_{\rm start}$ and
$\Delta t_{+}=t_{\rm end}-t_{\rm peak}$. The fitted coefficients
$\alpha$ and $\lambda_u$ are reported as model parameters but
are not interpreted as independently calibrated physical
quantities.
}
\end{table}